 \documentclass[sigconf]{acmart}
\AtBeginDocument{%
  }

\setcopyright{acmlicensed}
\copyrightyear{2026}
\acmYear{2026}
\acmDOI{XXXXXXX.XXXXXXX}
\acmConference[CHI26]{conference on Human Factors in Computing Systems }{April 13--17,
  2026}{Barcelona, ES}
\acmISBN{978-1-4503-XXXX-X/2018/06}

\usepackage{stmaryrd}
\begin{document}

\title{Human-Centered Evaluation of an LLM-Based Process Modeling
Copilot: A Mixed-Methods Study with Domain Experts}


\author{Chantale Lauer}
\email{chantale.lauer@uni-ssarland.de}
\orcid{0009-0001-6267-3265}
\affiliation{%
  \institution{Saarland University}
  \city{Saarbrücken}
  \state{Saarland}
  \country{Germany}
}
\affiliation{%
  \institution{German Research Institute for Artificial Intelligence}
  \city{Saarbrücken}
  \state{Saarland}
  \country{Germany}
}

\author{Peter Pfeiffer}
\email{peter.pfeiffer@5plus.ai}
\orcid{0000-0002-0224-4450}
\affiliation{%
  \institution{5Plus GmbH}
  \city{Würzburg}
  \state{Bavaria}
  \country{Germany}
}

\author{Nijat Mehdiyev}
\email{nijat.mehdiyev@dfki.de}
\orcid{0000-0001-7899-1017}
\affiliation{%
  \institution{German Research Institute for Artificial Intelligence}
  \city{Saarbrücken}
  \state{Saarland}
  \country{Germany}
}
\affiliation{%
  \institution{Saarland University}
  \city{Saarbrücken}
  \state{Saarland}
  \country{Germany}
}


\begin{abstract}
  Integrating Large Language Models (LLMs) into business process management tools promises to democratize Business Process Model and Notation (BPMN) modeling for non-experts. While automated frameworks assess syntactic and semantic quality, they miss human factors like trust, usability, and professional alignment. We conducted a mixed-methods evaluation of our proposed solution, an LLM-powered BPMN copilot, with five process modeling experts using focus groups and standardized questionnaires. Our findings reveal a critical tension between acceptable perceived usability (mean CUQ score: 67.2/100) and notably lower trust (mean score: 48.8\%), with reliability rated as the most critical concern (M=1.8/5). Furthermore, we identified output-quality issues, prompting difficulties, and a need for the LLM to ask more in-depth clarifying questions about the process. We envision five use cases ranging from domain-expert support to enterprise quality assurance.
  We demonstrate the necessity of human-centered evaluation complementing automated benchmarking for LLM modeling agents.
\end{abstract}


\begin{teaserfigure}
  \includegraphics[width=\textwidth]{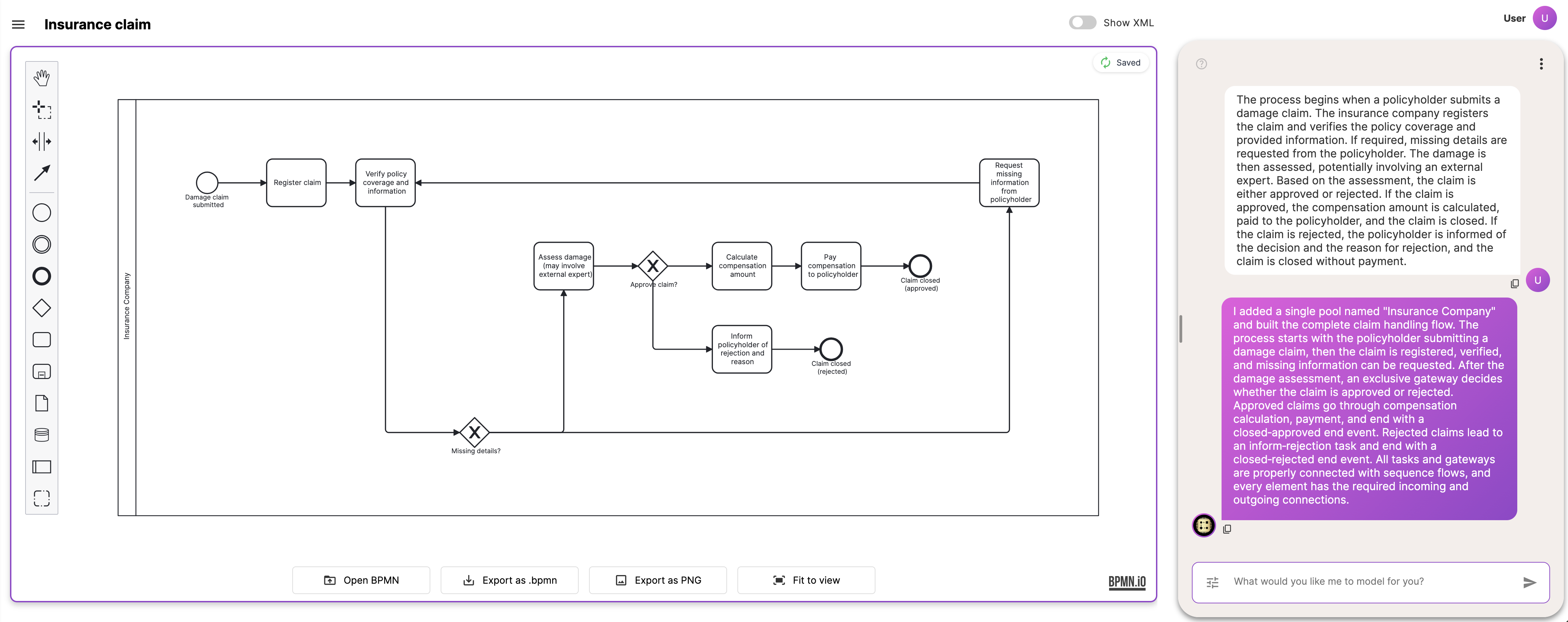}
  \caption{User interface of KICoPro with an example process.}
  \label{fig:KICoPro_UI}
\end{teaserfigure}


\maketitle
\section{Introduction}

Large Language Models (LLMs) are increasingly used in business process management (BPM), especially for generating Business Process Model and Notation (BPMN) models from natural language descriptions \cite{vanderaalst2013bpm,vidgof2023large,klievtsova2023conversational,kourani2024promoai}. Recent systems such as ProMoAI \cite{kourani2024promoai}, BPMN Chatbot \cite{kopke2024bpmnchatbot}, and Camunda BPMN Copilot \cite{drakopoulos2026llms} illustrate the potential of conversational process modeling. KICoPro\footnote{\url{https://www.kicopro.com/}} exemplifies such a conversational system that transforms textual process descriptions into BPMN models. This capability could broaden access to process modeling beyond dedicated experts \cite{vidgof2023large}.

However, existing evaluations have focused mainly on automated benchmarks of syntactic, semantic, and pragmatic quality \cite{kourani2025benchmark,drakopoulos2026llms}. While such metrics are important, they do not capture whether these systems are actually usable and trustworthy in practice. In particular, conversational BPMN generation may create a gulf of execution \cite{norman2013design}, where users must infer how to formulate prompts to achieve their modeling goals, and may increase cognitive load when they have to structure descriptions, decompose complex processes, and verify generated process models. These issues are especially relevant because BPMN models are human readable communication artifacts that must support collaborative understanding and review~\cite{omg2014bpmn}. As Liao and Varshney argue, human-centered evaluation is therefore needed to understand integration into work practices, trust calibration, and interaction breakdowns~\cite{liao2021humancentered}.

We address this gap through a focus group study with process modeling experts, complemented by standardized questionnaires measuring trust, usability, and task-specific performance. Our research questions are:

\begin{itemize}
    \item \textbf{RQ1:} How do process modeling experts perceive the usability and design of an LLM-based modeling copilot?
    \item \textbf{RQ2:} What are the strengths and weaknesses of LLM generated BPMN models from experts' perspectives?
    \item \textbf{RQ3:} What usage scenarios do experts envision for LLM based process modeling tools?
\end{itemize}

\section{Related Work}

Prior research shows that LLMs can support the generation and refinement of process models from textual descriptions. ProMoAI \cite{kourani2024promoai} demonstrated strong performance of GPT-4 in process model generation, error resolution, and feedback integration, while BPMN Chatbot \cite{kopke2024bpmnchatbot} showed that high model correctness can be achieved with substantially lower token usage. Overall, these studies highlight the potential of LLM based process modeling. But the evaluation is mainly based on quality metrics rather than user experience or trust. Commercial tools such as the Camunda BPMN Copilot~\cite{drakopoulos2026llms} and BPMN Assistant~\cite{licardo2025bpmnassistant} further reflect growing interest in LLM-assisted process modeling. Overall, the current work emphasizes technical performance over human factors in real-world use cases.

\subsection{Automated Evaluation Frameworks}

Several frameworks have been proposed for systematically evaluating LLM generated BPMN models. Kourani et al.~\cite{kourani2025benchmark} developed a comprehensive benchmark consisting of 20 diverse business processes, evaluating 16 state-of-the-art LLMs and revealing significant performance variations across LLMs. Recent work has also introduced structured evaluation frameworks assessing external quality across established BPM quality dimensions~\cite{drakopoulos2026llms,lauer2025evaluating, lauer2026assessingbusinessprocessmodeling}.

Drawing from ISO 9126 and systematic reviews of process model quality~\cite{sanchezgonzalez2010quality}, these frameworks typically evaluate clarity, correctness, and completeness. However, as Sànchez Gonzàlez et al.~\cite{sanchezgonzalez2010quality} argue, process model quality is inherently multidimensional, including pragmatic, human-centered aspects. However, those are not fully captured by automated metrics alone.

\subsection{Human Centered Evaluation of AI Systems}

The Human Centered Interaction (HCI) community has increasingly emphasized the need for human-centered evaluation of AI systems ~\cite{liao2021humancentered,amershi2019guidelines,yang2020reexamining}. 
Trust in AI systems has emerged as a critical factor in effective human AI collaboration~\cite{lee2004trust,devisser2020longitudinal}. Amershi et al.~\cite{amershi2019guidelines} proposed guidelines for human AI interaction that emphasize supporting user understanding, enabling appropriate trust, and facilitating effective correction, aspects that require human evaluation to assess. Scharowski et al.~\cite{scharowski2024trust} validated the Trust Scale for the AI Context (TAI), demonstrating its psychometric quality for measuring trust in AI applications, including chatbots.

For conversational AI, Holmes et al.~\cite{holmes2019cuq} developed the Chatbot Usability Questionnaire (CUQ), recognizing that traditional usability measures may not adequately capture the characteristics of conversational systems. The CUQ has been validated across multiple contexts~\cite{holmes2023validating} and adopted in studies of professional applications~\cite{familyhealthchatbot2024}.

\section{KICoPro}

KICoPro is a web-based conversational BPMN modeling tool that enables users to generate and iteratively refine BPMN models via natural language interaction. The system was developed as a research prototype to explore the potential of LLM-based assistance for process modeling tasks, with particular attention to supporting iterative refinements common in professional practice.

The system architecture combines a chat-based frontend, shown in \autoref{fig:KICoPro_UI}, with an LLM backend and BPMN rendering components. Users interact through a conversational interface where they can describe business processes in natural language (right-hand side in \autoref{fig:KICoPro_UI}), request modifications to generated models, ask questions about the current process model state, and request explanations of process modeling decisions. The system maintains a conversation history, enabling multi-turn interactions in which users can reference and refine previous outputs. The process models are displayed in the BPMN modeler (left-hand side in \autoref{fig:KICoPro_UI}), which also allows the users to modify the BPMN models.

KICoPro supports natural-language processing of domain-agnostic process descriptions, generates BPMN models that are visualizable within conversations, supports iterative refinement through targeted modifications, maintains persistent chat histories across sessions, enables BPMN file imports for extension, and offers compatibility with BPMN standard export.

At the time of evaluation, certain BPMN elements were not fully supported by the system's layout engine, including lanes (for representing organizational roles), message flows (for inter process communication), and some annotation types. These limitations were communicated to participants as part of the evaluation setup.
\section{Mixed-Method Evaluation}
We conducted a mixed-methods evaluation combining a focus group workshop with standardized questionnaires to capture both qualitative insights into expert users' experiences and quantitative measures of usability, trust, and task-specific performance. This methodological triangulation follows established practices in HCI research for evaluating complex interactive systems ~\cite{creswell2017mixed}, where neither purely qualitative nor purely quantitative approaches alone can capture the full picture of user experience. The qualitative component enables deep exploration of user perceptions, challenges, and envisioned use cases, while the quantitative component provides standardized measures that enable comparison with benchmarks and other systems.

\subsection{Participants}
Five process modeling experts participated in the evaluation ($n=5$), an overview is given in \autoref{tab:participants}. All participants worked extensively with BPMN models in their professional roles within the same organization, thereby representing a homogeneous sample of domain experts. Self-reported engagement with process models was mostly several days per week, indicating that process modeling constituted a core professional activity for all participants. Participant ages ranged across four decades (20s through 50s), representing diverse career stages. All participants reported prior experience with chatbot systems, with an average chatbot usage frequency of 2 to 3 days per week, indicating familiarity with conversational AI interfaces. 

\begin{table}[htbp]
\caption{Participant characteristics ($n=5$).}
\label{tab:participants}
\small
\begin{tabular}{lcccc}
\toprule
\textbf{ID} & \textbf{Age} & \textbf{BPMN Work} & \textbf{Chatbot} & \textbf{Chatbot Use} \\
 & \textbf{Group} & \textbf{(days/week)} & \textbf{Exp.} & \textbf{(days/week)} \\
\midrule
P1 & 20-30 & <1 & Yes & 2-3 \\
P2 & 20-30 & 2-3 & Yes & 2-3 \\
P3 & 20-30 & <1 & Yes & 2-3 \\
P4 & 31-40 & 4-5 & Yes & 3-5 \\
P5 & 51-60 & 4-5 & Yes & <1 \\
\bottomrule
\end{tabular}
\end{table}

Focusing on expert rather than novice users was a deliberate methodological choice aligned with our research questions. Experts identify subtle quality issues missed by novices, provide essential oversight for organizational adoption, and articulate workflow integration needs. 


\subsection{Procedure}
The evaluation was conducted in several phases, with a structured focus-group workshop lasting approximately three hours as the main component. It followed established guidelines for focus groups with expert participants~\cite{morgan1996focus,krueger2014focus}.

\textbf{Phase 1 - Kickoff (remote, 1 hour)} Participants received a brief kickoff session introducing the tool and its functionalities. They were then asked to share their expectations for a modeling chatbot without prior exposure to the prototype. This phase served dual purposes: orienting participants to the evaluation context and capturing baseline expectations prior to hands-on interaction.

\textbf{Phase 2 - Hands-on self exploration (2 weeks):} Following the kickoff, participants engaged in semi-structured, extensive hands-on exploration of our proposed solution. The participants were provided with two representative process descriptions, of varying complexity, to model, but were also encouraged to explore freely using their own process descriptions and modification requests. 

\textbf{Phase 3 - Focus Group (In-Person, 3 1/2 hours):} We conducted a semi-structured focus group where participants reflected on their experiences with our proposed solution, addressing interface usability, BPMN quality, limitations, improvements, and professional use cases. Discussions were documented through detailed observer notes and participant contributions on a shared whiteboard.

\textbf{Phase 4 - Questionnaires (30 minutes):} The participants completed standardized online questionnaires after the focus group, assessing usability, trust, and task-specific performance based on their overall hands-on experience.

\subsection{Instruments}
\subsubsection{Chatbot Usability Questionnaire (CUQ)}
We employed the 16-item Chatbot Usability Questionnaire (CUQ) \cite{holmes2019cuq} to evaluate perceived usability of the conversational BPMN copilot. As CUQ is specifically designed for chat-based interfaces, it overcomes limitations of traditional scales like System Usability Scale (SUS) \cite{brooke1996sus} by addressing conversational characteristics across five dimensions: personality and engagement, onboarding and welcome, navigation and ease of use, understanding and communication, and error handling and overall assessment. Participants responded on a 5-point Likert scale (1 = strongly disagree to 5 = strongly agree). Eight negative items were reverse-scored before being aggregated into a 0–100 score, with higher values indicating superior usability. Through validation studies \cite{holmes2023validating}, a benchmark mean of 68/100 was established.

\subsubsection{Trust Assessment}
Trust was measured using an 8-item scale adapted from the Trust Scale for the AI Context (TAI)~\cite{scharowski2024trust,hoffman2023tai}, which has been validated for measuring trust in AI-based chatbot systems with strong psychometric properties. The scale captures multiple dimensions of trust in AI systems: confidence in functionality, predictability, reliability, security, efficiency, suspicion, comparative competence, and decision support. Each item was rated on a 5-point Likert scale, with the total score converted to a percentage scale (range: 0-100\%) for interpretability.

\subsubsection{Tool-Specific Quality Assessment}
We developed an eight-item questionnaire to assess BPMN model quality and task-specific capabilities not captured by generic measures. The items covered understanding of textual descriptions, representation of key activities, correct sequencing, structural clarity, implementation of modifications, process explanations, professional suitability, and the level of detail. The responses, given on a five-point Likert scale, were converted to percentages. Additionally, two open-ended questions captured positive and negative use cases: "I like using the tool for..." and "The tool did not achieve good results in the following use cases:...".

\subsection{Data Analysis}
All quantitative questionnaire data were analyzed descriptively (means, ranges, distributions, and standard deviation). Given the small sample size ($n=5$), we report descriptive statistics following study recommendations ~\cite{sauro2016quantifying}. The small sample precludes claims about statistical significance but enables identification of patterns.

Qualitative data from the focus group, workshop documentation, and open-ended questionnaire responses were analyzed using thematic analysis following Braun and Clarke's six-phase approach \cite{Braun2006Qualitative}. The initial coding captured discrete observations, which were iteratively clustered into candidate themes based on conceptual similarity. Those themes were refined against the full dataset for coherence and organized around the study's three research questions, highlighting both common patterns and distinctive perspectives.

\section{Results}

\subsection{Qualitative Results}
Thematic analysis of focus group discussions, workshop documentation, and open-ended questionnaire responses yielded seven major themes organized around the three research questions, revealing seven major themes, as well as five potential use cases.


\noindent \textbf{RQ1: Usability and Design Perceptions} 

\textit{Theme 1: Intuitive Interface, Opaque Prompting.} The interface was praised as simple and intuitive, with familiar chat-based operations (text input, process model viewing, and new conversations) posing no difficulties. 
However, a prompting paradox revealed a gulf of execution: users understood the overall goal of generating BPMN models from text, but were unsure how to formulate prompts that would reliably produce useful results. They struggled with input formulation, e.g., balancing detail vs. conciseness, structuring multi part processes, and including domain context, which hindered the adoption. Error handling exacerbated this, as participants lacked remediation guidance for layout errors or suboptimal outputs.

\textit{Theme 2: Response Latency as Workflow Barrier.} The process modeling time was perceived as overly long, disrupting the iterative workflow essential to professional process modeling. Participants reported that input output latency impeded rapid iteration for refinement and exploration, disrupting the process modeler’s cognitive rhythm. \\

\noindent \textbf{RQ2: Strengths and Weaknesses of LLM-Generated BPMNs}

\textit{Theme 3: Varying Output Quality} Participants report that the quality of generated BPMN models is sometimes insufficient. In particular, longer process descriptions tend to yield lower-quality BPMN models that capture only a subset of the described process.

\textit{Theme 4: Chunking as Emergent Coping Strategy.} 
Participants discovered chunking as a coping strategy to improve output quality for long descriptions, with iterative, piecewise prompting outperforming comprehensive inputs. However, this strategy increased cognitive load, as users had to segment their mental model of the process into manageable fragments, maintain coherence across iterations, and mentally reassemble the resulting model.

\textit{Theme 5: Imprecise Modifications.} Modification requests proved unreliable due to incorrect implementations. Key issues included unsupported BPMN elements (lanes, annotations) without notification, with participants stressing that complete palette coverage is needed. Additionally, modification requests often trigger issues such as insertions with unexpected connections or unintended modifications to previously unchanged parts.

\textit{Theme 6: Absent Clarification Dialogue.} It was noted that the LLM failed to ask clarifying questions for ambiguous inputs. Professional process modeling requires an iterative dialogue to resolve descriptive gaps, a capability absent in the system. This lack of clarifying interaction leads the LLM to generate complete outputs from incomplete inputs via implicit rather than explicit assumptions.

\textit{Theme 7: Convention Violations.} Participants observed that process modeling conventions were not followed, including both BPMN 2.0 standards (e.g., a task has one incoming and one outgoing sequence flow) as well as organization-specific guidelines (e.g., labels need to follow a specific pattern) intended to ensure consistency across an enterprise process portfolio. They expected the tool to be configurable with enterprise conventions and capable of performing quality assurance against them. \\

\noindent \textbf{RQ3: Envisioned Use Cases} \\
Analysis of open-ended responses and focus group discussions revealed five distinct use cases that the participants envisioned for LLM-based process modeling tools. 

\textit{Use Case 1: Support for Domain Experts (Non-Modelers).} The system aids BPMN-inexperienced domain experts by generating diagrams from natural language, enabling specialist refinement or direct documentation, also addressing the "blank page problem" by providing an initial structure. However, it was stressed that non-experts need higher output quality, as they cannot readily detect errors, unlike skilled process modelers.

\textit{Use Case 2: Quality Assurance Bot.} The copilot could validate process models against BPMN standards and organizational conventions, while automatically correcting identified issues. 

\textit{Use Case 3: Process Creation from Visual Inputs.} A key workshop requirement was image-to-BPMN conversion, covering photographs of hand-drawn sketches. This addresses the common need to formalize informal process model sketches. 

\textit{Use Case 4: Enterprise-Integrated Local Deployment.} A local LLM variant was envisioned, which is trained on the organization's process portfolio, enabling the recognition of existing processes, pattern-based suggestions, enterprise-wide consistency checks, and subprocess reuse. 

\textit{Use Case 5: Process Optimization and Support.} The system should support process improvement by identifying optimization potential, suggesting enhancements, and asking probing questions. This extends beyond modeling to active process analysis, leveraging LLM reasoning to detect inefficiencies human modelers might miss. \\

\textit{Negative Use Cases.} Participants also clearly identified contexts in which the current system performed poorly, thereby highlighting its present limitations. Long and complex process descriptions often led to unsatisfactory results (P1, P3, P4), and reconstructing already known processes from memory also proved challenging (P5). Moreover, for experts working on complex processes, the tool was perceived as slowing work down rather than accelerating it (P2). These negative cases help bound expectations for current capabilities and identify priority areas for improvement.


\subsection{Quantitative Results}
This section summarizes the results from the questionnaires concerning usability, trust, and task-specific quality. Table~\ref{tab:scores-summary} summarizes the scores across all assessment dimensions.
\begin{table}[htbp]
\caption{Summary of quantitative assessment scores.}
\label{tab:scores-summary}
\small
\begin{tabular}{lccccc}
\toprule
\textbf{Measure} & \textbf{Mean} & \textbf{SD} & \textbf{Min} & \textbf{Max} & \textbf{Ref.} \\
\midrule
Usability (CUQ) & 67.2 & 7.1 & 56.25 & 73.44 & 68.0$^\dagger$ \\
Trust & 48.8 & 10.5 & 31.25 & 59.38 & 60.0$^\ddagger$ \\
Task Quality & 54.4 & 17.8 & 25.0 & 65.63 & -- \\
\bottomrule
\multicolumn{6}{l}{$^\dagger$CUQ benchmark. $^\ddagger$Suggested acceptable threshold.}
\end{tabular}
\end{table}

\subsubsection{Chatbot Usability (CUQ)}

The mean CUQ score across all participants was 67.2 ($SD = 7.1$), falling marginally below the established benchmark of 68 for chatbot usability. Individual scores ranged from 56.25 to 73.44, indicating moderate variability in usability perceptions across participants. The median score (70.31) slightly exceeded the mean, suggesting one lower outlier.
\begin{table}[htbp]
\caption{Chatbot Usability Questionnaire (CUQ) item-Level results.}
\label{tab:cuq-items}
\small
\begin{tabular}{clcccc}
\toprule
\textbf{Item} & \textbf{Statement} & \textbf{Mean} & \textbf{SD} & \textbf{Int.} & \textbf{Critical} \\
\midrule
Q1 (+) & Realistic personality & 3.0 & 1.00 & $\circ$ &   \\
Q2 (--) & Robotic behavior & 2.4 & 0.89 & + &  \\
Q3 (+) & Welcoming onboarding & \textbf{4.4} & 0.55 & + &  \\
Q4 (--) & Unfriendly behavior & \textbf{1.2} & 0.45 & ++ &  \\
Q5 (+) & Clear purpose explanation & 2.2 & 0.45 & - &  ${\;\lightning\;}$ \\
Q6 (--) & No purpose explanation & 3.2 & 1.30 & $\circ$ &  \\
Q7 (+) & Easy to use & \textbf{4.0} & 0.71 & + &  \\
Q8 (--) & Confusing to use & 2.4 & 1.34 & + &   \\
Q9 (+) & User input understanding & 3.2 & 1.30 & $\circ$ &  \\
Q10 (--) & Input recognition failure & 2.6 & 0.89 & $\circ$&   \\
Q11 (+) & Useful responses & \textbf{3.8} & 0.45 & + &   \\
Q12 (--) & Irrelevant responses & \textbf{1.4} & 0.55 & ++ &  \\
Q13 (+) & Good error handling & 3.4 & 1.14 & $\circ$ &  \\
Q14 (--) & Unable to handle errors & 1.8 & 1.30 & - & ${\;\lightning\;}$\\
Q15 (+) & Overall ease of use & \textbf{4.0} & 1.22 & +&   \\
Q16 (--) & Complex usage & 2.0 & 0.71 & +&   \\
\bottomrule
\multicolumn{6}{l}{\footnotesize Scale: 1 = strongly disagree, 5 = strongly agree. Item: (+) = positive; (-) = negative. } \\
\multicolumn{6}{l}{\footnotesize Int.: ++ = highly positive; + = positive; $\circ$= neutral; - = negative; -- = highly negative; } \\
\end{tabular}
\end{table}

Analysis of individual CUQ items, shown in \autoref{tab:cuq-items}, revealed important patterns in usability perceptions. The highest-rated items related to the initial user experience and basic interaction mechanics. Participants found the chatbot welcoming at first setup (Q3: $M = 4.4$, $SD = 0.55$), easy to use (Q7: $M = 4.0$, $SD = 0.71$; Q15: $M = 4.0$, $SD = 1.22$), and not unfriendly (Q4: $M = 1.2$, $SD = 0.45$) or excessively complex (Q16: $M = 2.0$, $SD = 0.71$). Responses were generally perceived as useful, appropriate, and informative (Q11: $M = 3.8$, $SD = 0.45$), and the system was not seen as producing irrelevant responses (Q12: $M = 1.4$, $SD = 0.55$).

However, participants showed uncertainty about the system's transparency: the item "clear purpose explanation" scored below neutral (Q5: $M=2.2$, $SD=0.45$), while "no purpose explanation" exhibited high variability (Q6: $M=3.2$, $SD=1.30$), indicating disagreement. This suggests that despite ease of use, users lacked clarity on system capabilities, which aligns with qualitative findings. The understanding of user input showed mixed perceptions (Q9: $M=3.2$, $SD=1.30$), with uncertainty about input recognition failures (Q10: $M=2.6$, $SD=0.89$). This high variability suggests that comprehension depends on input characteristics or strategies. Also, error handling received moderate ratings (Q13: $M=3.4$, $SD=1.14$; Q14: $M=1.8$, $SD=1.30$), again with notable variability.

\subsubsection{Trust Assessment}

The mean trust score, computed as the average of participants’ aggregated item scores, was 48.8\% ($SD = 10.5\%$), indicating moderate trust in the system. However, this score falls notably below a suggested threshold of 60\% for acceptable trust and substantially below the usability score, revealing a usability-trust gap. Individual scores ranged from 31.25\% to 59.38\%, with participant P5 showing notably lower trust than others.


\begin{table}[htbp]
\caption{Trust scale item-Level results.}
\label{tab:trust-items}
\small
\begin{tabular}{clcclc}
\toprule
\textbf{Item} & \textbf{Statement}& \textbf{Mean} & \textbf{SD} & \textbf{Int} &  \textbf{Critical}\\
\midrule
Q1 (+) & Functions as intended & 3.6 & 0.55 & + &\\
Q2 (+) &Predictable results & 2.4 & 0.55 & $\circ$ &\\
Q3 (+) &\textbf{Correctness reliability} & \textbf{1.8} & \textbf{0.45} & - & ${\;\lightning\;}$ \ \\
Q4 (+) &Safe to rely on & 2.4 & 0.55 & $\circ$ &  \\
Q5 (+) &Operational efficiency & 3.2 & 0.84 & $\circ$ &\\
Q6 (-) &System suspicion (rev.) & 3.8 & 1.10 & + & \\
Q7 (+) &Outperforms novice human & 3.8 & 0.84 & + & \\
Q8 (+) &Suitable for decision-making & 2.6 & 0.89 & $\circ$ & \\
\bottomrule
\multicolumn{6}{l}{\footnotesize Scale: 1 = strongly disagree, 5 = strongly agree. Item: (+) = positive; (-) = negative. } \\
\multicolumn{6}{l}{\footnotesize Int.: ++ = highly positive; + = positive; $\circ$= neutral; - = negative; -- = highly negative; } \\
\end{tabular}
\end{table}

Item-level analysis in \autoref{tab:trust-items} revealed critical patterns in trust formation. The participants expressed moderate confidence that the system functions well overall (Q1: $M = 3.6$, $SD = 0.55$) and believed the system could perform the task better than an inexperienced human user (Q7: $M = 3.8$, $SD = 0.84$).

However, critical trust dimensions received notably lower ratings. Reliability received the lowest rating (Q3: $M=1.8$, $SD=0.45$), indicating that the participants could not consistently rely on correct outputs. As this showed the lowest variability, it suggests consensus. Predictability was also rated low (Q2: $M = 2.4$, $SD = 0.55$), as was the sense of security when relying on the system (Q4: $M = 2.4$, $SD = 0.55$). The system's suitability for decision-making received moderate ratings (Q8: $M = 2.6$, $SD = 0.89$).

Notably, participants showed low suspicion of the system (Q6: $M=3.8$, $SD=1.10$) and moderate efficiency ratings (Q5: $M=3.2$, $SD=0.84$). This indicates no distrust of malicious intent but rather concerns about output reliability and correctness, revealing a nuanced trust profile where the system is seen as well-intentioned but inconsistent.

\subsubsection{Tool-Specific Quality Assessment}
The mean score for the tool-specific quality assessment was 54.4\% ($SD = 17.8\%$), indicating moderate perceived quality of generated BPMN models. Individual scores ranged widely from 25.0\% to 65.63\%, with participant P5 rating quality substantially lower than others. As it was the same participant who showed lowest trust and usability scores, it suggests consistency in that individual's negative experience.

\begin{table}[htbp]
\caption{Tool-specific quality assessment results.}
\label{tab:quality-items}
\small
\begin{tabular}{clcclc}
\toprule
\textbf{Item} & \textbf{Statement} & \textbf{Mean} & \textbf{SD} & \textbf{Int.} & \textbf{Critical} \\
\midrule
Q1 (+) & Text understanding & 2.8 & 1.30 & - & ${\;\lightning\;}$ \\
Q2 (+) &Process representation & 3.0 & 0.71 & $\circ$&  \\
Q3 (+) &Correct flow order & 3.4 & 1.34 & - & ${\;\lightning\;}$\\
Q4 (+) &Structural clarity & 3.2 & 0.84 & $\circ$ & \\
Q5 (+) &Modification handling & 3.2 & 1.30 & - & ${\;\lightning\;}$\\
Q6 (+) &Description correctness & 3.4 & 0.55 & $\circ$  &\\
Q7 (+) &Application suitability & 3.4 & 1.34 & - & ${\;\lightning\;}$\\
Q8 (+) &\textbf{Level of detail} & \textbf{3.8} & \textbf{0.45} & + & \\
\bottomrule
\multicolumn{6}{l}{\footnotesize Scale: 1 = strongly disagree, 5 = strongly agree. Item: (+) = positive; (-) = negative. } \\
\multicolumn{6}{l}{\footnotesize Int.: ++ = highly positive; + = positive; $\circ$= neutral; - = negative; -- = highly negative; } \\
\end{tabular}
\end{table}

The item-level analysis (Table~\ref{tab:quality-items}) revealed that the level of detail received the highest rating (Q8: $M = 3.8$, $SD = 0.45$), suggesting that generated process models captured appropriate granularity. The system's ability to provide correct descriptions of modeled processes was also rated moderately well (Q6: $M = 3.4$, $SD = 0.55$), as were activity flow and ordering (Q3: $M = 3.4$, $SD = 1.34$) and overall application appropriate quality (Q7: $M = 3.4$, $SD = 1.34$).

Text understanding received low ratings with high variability (Q1: $M=2.8$, $SD=1.30$), indicating inconsistent input comprehension. This suggests comprehension depends on input characteristics or participant discovered interaction strategies. Also, modification handling showed similar patterns (Q5: $M = 3.2$, $SD = 1.30$), indicating variable success for change requests. 


\section{Discussion}

Our results show that human-centered evaluation reveals system quality aspects beyond automated metrics~\cite{kourani2025benchmark,drakopoulos2026llms}. Specifically, our study uncovered a usability trust gap: a pleasant interaction does not guarantee confidence in output reliability for professional use. This observation aligns with trust calibration research distinguishing affective responses from cognitive assessments of system capability~\cite{lee2004trust}. These findings, together with the other results reported in this study, motivate design implications that can only be identified through human-centered evaluation. Accordingly, future systems should be assessed using both automated benchmarks for scalability and human evaluation to surface interaction breakdowns, trust needs, and organizational fit that inform design.

\subsection{Design Implications}
\textbf{Prompting Guidance.} The prompting paradox indicates a need for explicit input formulation support by making system expectations transparent, thus reducing the gulf of execution~\cite{norman2013design}. This could include example conversations, templates for common process types and documentation of what input features correlate with good output quality. 

\textbf{Progressive Disclosure and Chunking Support.} The success of chunking suggests designing systems that can incrementally construct a process model from a large process description, rather than requiring users to chunk the input themselves, thereby reducing users' cognitive load and improving outcomes. This could include staged information prompts, piece-by-piece building with intermediate validation, and state visualization. 

\textbf{Proactive Clarification.} Future systems should detect input ambiguities and ask clarifying questions before generating output. 

\textbf{Convention Configuration.} Organizational conventions necessitate customization capabilities for professional deployment. Systems should encode conventions, validate outputs against them, and learn from existing process portfolios.

\textbf{Confidence Communication.} Given reliability concerns, systems should communicate output confidence, highlight uncertainties or likely errors, and direct review to problematic elements. This transparency supports appropriate trust calibration~\cite{lee2004trust}.

\textbf{Latency Management.} Response time concerns suggest the need for progress indication during generation, streaming output where feasible, and interaction design that accommodates processing delays without disrupting cognitive flow.

\subsection{Limitations}
Several limitations constrain our findings' generalizability. The small ($n=5$), single-organization expert sample limits applicability to novices and other contexts. Moreover, our expert focus limits conclusions about democratization potential. The evaluation captured a snapshot of an evolving system, in which technical limitations (e.g., missing lanes) have since been addressed. Prior chatbot experience may have shaped participant expectations differently than for first-time users. Also, the workshop format may not reflect extended use patterns. Despite these, our qualitative insights complement automated evaluations and inform larger, more diverse studies.

\section{Conclusion}
We conducted a mixed methods evaluation of our LLM-based conversational BPMN modeling system, with five domain experts using a focus group workshop and questionnaires. Our quantitative analysis revealed a usability trust gap: although usability scores approached established benchmarks, trust in output reliability lagged, with reliability emerging as the primary concern. The qualitative analysis identified seven key themes in experts' experiences, including the prompting paradox (knowing the tool's function but not its effective use), output quality issues, particularly for long textual descriptions, and an absent clarifying dialogue. Participants envisioned use cases, including initial drafts for domain experts, quality assurance of existing models, and image-to-BPMN support.

Our work shows that human-centered evaluation reveals critical LLM toolaspects of LLM tool quality, such as trust calibration, interaction challenges, coping strategies, and alignment with professional practice automated benchmarks. Therefore, we argue, that a comprehensive evaluation must consist of an automated technical assessment combined with a human-centered practical investigation. Understanding experts’ experiences, trust, and adaptation is essential for practical adoption, as these tools are intended for use in professional domains.

\begin{acks}
Parts of this work were conducted within the projects KICoPro (FKZ: 01IS24053C) and EINHORN (FKZ: 01IS24048C), funded by the Federal Ministry of Research, Technology and Space (BMFTR).
\end{acks}

\bibliographystyle{ACM-Reference-Format}
\bibliography{references}

@article{vanderaalst2013bpm,
  author    = {van der Aalst, Wil M. P.},
  title     = {Business Process Management: A Comprehensive Survey},
  journal   = {ISRN Software Engineering},
  volume    = {2013},
  pages     = {1--37},
  year      = {2013},
  doi       = {10.1155/2013/507984}
}

@inproceedings{vidgof2023large,
  title={Large language models for business process management: Opportunities and challenges},
  author={Vidgof, Maxim and Bachhofner, Stefan and Mendling, Jan},
  booktitle={International conference on business process management},
  pages={107--123},
  year={2023},
  organization={Springer}
}

@inproceedings{klievtsova2023conversational,
  author    = {Klievtsova, Nataliia and Bezin, Janik and Kampik, Timotheus and Mangler, Jürgen and Rinderle-Ma, Stefanie},
  title     = {Conversational Process Modelling: State of the Art, Applications, and Implications in Practice},
  booktitle = {Business Process Management Forum},
  publisher = {Springer},
  year      = {2023},
  pages     = {319--336},
  doi       = {10.1007/978-3-031-41623-1_19}
}

@article{kourani2024promoai,
  author    = {Kourani, Humam and Berti, Alessandro and van der Aalst, Wil M. P.},
  title     = {Process Modeling With Large Language Models},
  journal   = {arXiv preprint arXiv:2403.07541},
  year      = {2024}
}

@inproceedings{kopke2024bpmnchatbot,
  author    = {K{\"o}pke, Julius and Safan, Aya},
  title     = {Introducing the {BPMN-Chatbot} for Efficient {LLM-Based} Process Modeling},
  booktitle = {Proceedings of the BPM 2024 Demos \& Resources Forum},
  series    = {CEUR Workshop Proceedings},
  volume    = {3758},
  year      = {2024},
  pages     = {86--90}
}

@article{drakopoulos2026llms,
  title={Do LLMs Speak BPMN? An Evaluation of Their Process Modeling Capabilities Based on Quality Measures},
  author={Drakopoulos, Panagiotis and Malousoudis, Panagiotis and Nousias, Nikolaos and Tsakalidis, George and Vergidis, Kostas},
  journal={Computation},
  volume={14},
  number={1},
  pages={10},
  year={2026},
doi = {10.3390/computation14010010},
  publisher={MDPI}
}

@article{kourani2025benchmark,
  author    = {Kourani, Humam and Berti, Alessandro and Schuster, Daniel and Van der Aalst, Wil M.P.},
  title     = {Evaluating Large Language Models on Business Process Modeling: Framework, Benchmark, and Self-Improvement Analysis},
  journal   = {Software and Systems Modeling},
  year      = {2025},
  doi       = {10.1007/s10270-025-01318-w}
}

@inproceedings{lauer2025evaluating,
  author    = {Lauer, Chantale and Pfeiffer, Peter and Rombach, Alexander and Mehdiyev, Nijat},
  title     = {Conversational Business Process Modeling using LLMs: Initial Results and Challenges},
  booktitle = {EMISA 2025 Proceedings},
  publisher = {GI},
  year      = {2025}
}

@article{licardo2025bpmnassistant,
  author    = {Licardo, Josip Tomo and Tankovic, Nikola and Etinger, Darko},
  title     = {{BPMN Assistant}: An {LLM-Based} Approach to Business Process Modeling},
  journal   = {arXiv preprint arXiv:2509.24592},
  year      = {2025}
}

@techreport{omg2014bpmn,
  author      = {{Object Management Group}},
  title       = {Business Process Model and Notation ({BPMN}), Version 2.0.2},
  institution = {OMG},
  year        = {2014},
  url         = {https://www.omg.org/spec/BPMN/2.0.2}
}

@article{sanchezgonzalez2010quality,
  author    = {S{\'a}nchez-Gonz{\'a}lez, Laura and Rubio, F{\'e}lix Garc{\'i}a and Gonz{\'a}lez, Francisco Ruiz and Velthuis, Mario Piattini},
  title     = {Measurement in business processes: a systematic review},
  journal   = {Business Process Management Journal},
  year      = {2010},
  pages     = {114--134}, 
  doi       = {https://doi.org/10.1108/14637151011017976}
}

@article{liao2021humancentered,
  author    = {Liao, Q. Vera and Varshney, Kush R.},
  title     = {Human-Centered Explainable {AI (XAI)}: From Algorithms to User Experiences},
  journal   = {arXiv preprint arXiv:2110.10790},
  year      = {2021}
}

@inproceedings{amershi2019guidelines,
  author    = {Amershi, Saleema and Weld, Dan and Vorvoreanu, Mihaela and Fourney, Adam and Nushi, Besmira and Collisson, Penny and Suh, Jina and Iqbal, Shamsi and Bennett, Paul N. and Inkpen, Kori and Teevan, Jaime and Kikin-Gil, Ruth and Horvitz, Eric},
  title     = {Guidelines for Human-{AI} Interaction},
  booktitle = {Proceedings of the 2019 CHI Conference on Human Factors in Computing Systems},
  series    = {CHI '19},
  year      = {2019},
  pages     = {1--13},
  doi       = {10.1145/3290605.3300233}
}

@inproceedings{yang2020reexamining,
  author    = {Yang, Qian and Steinfeld, Aaron and Ros{\'e}, Carolyn and Zimmerman, John},
  title     = {Re-Examining Whether, Why, and How Human-{AI} Interaction Is Uniquely Difficult to Design},
  booktitle = {Proceedings of the 2020 CHI Conference on Human Factors in Computing Systems},
  series    = {CHI '20},
  year      = {2020},
  pages     = {1--13},
  doi       = {10.1145/3313831.3376301}
}

@article{lee2004trust,
  author    = {Lee, John D. and See, Katrina A.},
  title     = {Trust in Automation: Designing for Appropriate Reliance},
  journal   = {Human Factors},
  volume    = {46},
  number    = {1},
  pages     = {50--80},
  year      = {2004},
  doi       = {10.1518/hfes.46.1.50_30392}
}

@article{devisser2020longitudinal,
  author    = {de Visser, Ewart J. and Peeters, Marieke M. M. and Jung, Malte F. and Kohn, Spencer and Shaw, Tyler H. and Pak, Richard and Neerincx, Mark A.},
  title     = {Towards a Theory of Longitudinal Trust Calibration in Human--Robot Teams},
  journal   = {International Journal of Social Robotics},
  volume    = {12},
  pages     = {459--478},
  year      = {2020},
  doi       = {10.1007/s12369-019-00596-x}
}

@inproceedings{scharowski2024trust,
  title={To Trust or Distrust AI: A Questionnaire Validation Study},
  author={Scharowski, Nicolas and Perrig, Sebastian A.C. and von Felten, Nick and Aeschbach, Lena Fanya and Opwis, Klaus and Wintersberger, Philipp and Br{\"u}hlmann, Florian},
  booktitle={Proceedings of the 2025 ACM Conference on Fairness, Accountability, and Transparency},
  pages={361--374},
  year={2025}
}

@book{hoffman2023tai,
  author    = {Hoffman, Robert R. and others},
  title     = {Metrics for Explainable {AI}},
  publisher = {CRC Press},
  year      = {2023},
  chapter   = {Trust Scale for the AI Context (TAI)}
}

@book{norman2013design,
  author    = {Norman, Don},
  title     = {The Design of Everyday Things: Revised and Expanded Edition},
  publisher = {Basic Books},
  location  = {London}, 
  year      = {2013}
}

@misc{lauer2026assessingbusinessprocessmodeling,
      title={Assessing the Business Process Modeling Competences of Large Language Models}, 
      author={Chantale Lauer and Peter Pfeiffer and Alexander Rombach and Nijat Mehdiyev},
      year={2026},
      eprint={2601.21787},
      archivePrefix={arXiv},
      primaryClass={cs.SE},
      url={https://arxiv.org/abs/2601.21787}, 
}

@inproceedings{holmes2019cuq,
  author    = {Holmes, Samuel and Moorhead, Anne and Bond, Raymond and Zheng, Huiru and Coates, Vivien and McTear, Michael},
  title     = {Usability Testing of a Healthcare Chatbot: Can We Use Conventional Methods to Assess Conversational User Interfaces?},
  booktitle = {Proceedings of the 31st European Conference on Cognitive Ergonomics},
  series    = {ECCE '19},
  year      = {2019},
  pages     = {207--214},
  doi       = {10.1145/3335082.3335094}
}

@inproceedings{holmes2023validating,
  author    = {Holmes, Samuel and Bond, Raymond and Moorhead, Anne and Zheng, Jane and Coates, Vivien and McTear, Michael},
  title     = {Towards Validating a Chatbot Usability Scale},
  booktitle = {Design, User Experience, and Usability. HCII 2023},
  series    = {Lecture Notes in Computer Science},
  volume    = {14033},
  publisher = {Springer},
  year      = {2023},
  pages     = {321--339},
  doi       = {10.1007/978-3-031-35708-4_24}
}

@article{familyhealthchatbot2024,
  author    = {{Nguyen, Michelle Hoang and Sedoc, Jo{\~a}o and Taylor, Casey Overby}},
  title     = {Usability, Engagement, and Report Usefulness of Chatbot-Based Family Health History Data Collection: Mixed Methods Analysis},
  journal   = {Journal of Medical Internet Research},
  volume    = {26},
  pages     = {e55164},
  year      = {2024},
  doi       = {10.2196/55164}
}

@article{brooke1996sus,
  author    = {Brooke, John},
  title     = {{SUS}: A `Quick and Dirty' Usability Scale},
  journal   = {Usability Evaluation in Industry},
  year      = {1996},
  pages     = {189--194}
}

@article{morgan1996focus,
  author    = {Morgan, David L.},
  title     = {Focus Groups},
  journal   = {Annual Review of Sociology},
  volume    = {22},
  number    = {1},
  pages     = {129--152},
  year      = {1996},
  doi       = {10.1146/annurev.soc.22.1.129}
}

@book{krueger2014focus,
  author    = {Krueger, Richard A. and Casey, Mary Anne},
  title     = {Focus Groups: A Practical Guide for Applied Research},
  publisher = {SAGE},
  edition   = {5th},
  year      = {2014}
}

@book{creswell2017mixed,
  author    = {Creswell, John W. and Clark, Vicki L. Plano},
  title     = {Designing and Conducting Mixed Methods Research},
  publisher = {SAGE},
  edition   = {3rd},
  year      = {2017}
}

@book{sauro2016quantifying,
author = {Lewis, James R. and Sauro, Jeff},
year = {2016},
month = {08},
pages = {},
title = {Quantifying the User Experience (2nd ed.)},
isbn = {978-0128023082}
}

@article{Braun2006Qualitative,
   author = {Virginia Braun and Victoria Clarke},
   doi = {10.1191/1478088706QP063OA},
   issn = {14780887},
   issue = {2},
   journal = {Qualitative Research in Psychology},
   pages = {77-101},
   publisher = {Taylor \& Francis Group},
   title = {Using thematic analysis in psychology},
   volume = {3},
   url = {https://www.tandfonline.com/doi/abs/10.1191/1478088706qp063oa},
   year = {2006}
}

\end{document}